\documentclass[onecolumn,fleqn,nobibnotes]{revtex4}

\usepackage{amsmath}
\usepackage{graphicx}
\usepackage{float}
\usepackage{subfigure}
\usepackage{wrapfig}
\usepackage{multirow}
\usepackage{tabularx}
\usepackage{array}
\usepackage{url}
\usepackage{tikz}
\usetikzlibrary{positioning}

\def\lsim{\raise0.3ex\hbox{$<$\kern-0.75em\raise-1.1ex\hbox{$\sim$}}}
\def\gsim{\raise0.3ex\hbox{$>$\kern-0.75em\raise-1.1ex\hbox{$\sim$}}}

\def\beq{\begin{equation}}
\def\eeq{\end{equation}}
\def\beqa{\begin{eqnarray}}
\def\eeqa{\end{eqnarray}}

\newcommand{\la}{\langle}
\newcommand{\ra}{\rangle}

\def\gappeq{\mathrel{\rlap {\raise.5ex\hbox{$>$}}
{\lower.5ex\hbox{$\sim$}}}}

\def\lappeq{\mathrel{\rlap{\raise.5ex\hbox{$<$}}
{\lower.5ex\hbox{$\sim$}}}}

\def\Toprel#1\over#2{\mathrel{\mathop{#2}\limits^{#1}}}

\begin{document}

\title{Magnetic field in relativistic heavy ion collisions: testing the  
classical approximation}
\author{I. Danhoni$^1$ and F. S. Navarra$^1$}
\affiliation{$^1$Instituto de F\'{\i}sica, Universidade de S\~{a}o Paulo, 
 Rua do Mat\~ao, 1371, CEP 05508-090,  S\~{a}o Paulo, SP, Brazil\\
}
\begin{abstract}
It is believed that in non-central relativistic heavy ion collisions a  
very strong
magnetic field is formed. There are several studies of the effects of
this field, where $\vec{B}$ is calculated with the expressions of classical
electrodynamics. A quantum field may be approximated by a classical one when
the number of field quanta in each field mode is sufficiently high. This may
happen if the field sources are intense enough. In  heavy ion physics the
validity of the classical treatment was not investigated. 
In this work we propose a test of the quality of the classical approximation.
We calculate
an observable quantity using the classical magnetic field  and also using
photons as input. If the results of both approaches coincide, this will be an
indication that the classical approximation is valid. 
More precisely, we focus on  the process in which a nucleon is converted into a
delta resonance, which then decays into another nucleon and a pion, i.e.,
$ N \to \Delta \to N' \pi$. In ultra-peripheral relativistic heavy
ion collisions this conversion can be induced by the classical magnetic field 
of one the ions acting on the other ion. Alternatively, we can replace the   
classical magnetic field by a flux of equivalent photons, which are absorbed   
by the target nucleons. We calculate the cross sections in these two independent
ways and find that they differ from each other by $\simeq 10$ \% in the
considered collision energy range. This suggests that the two formalisms are
equivalent and that the classical approximation for the magnetic field is
reasonable. 

\end{abstract}
\maketitle

\section{Introduction}

It has been often said that in relativistic heavy ion collisions we produce   
the strongest magnetic field of the universe \cite{skokov,voro,muller2}. This
field is so intense because the charge density is large, 
because the speed of the source is very close to the speed of light and also
because we probe it at extremely small distances (a few fermi) from the
source.

There has been a search for observable effects of this strong field
\cite{hattori}. The first and most famous is the Chiral Magnetic Effect (CME) 
\cite{cme}. A natural place to look for this field and its effects is
in ultra-peripheral relativistic heavy ion collisions (UPC's), in which the
two nuclei do not overlap \cite{upc}.  
Since there is no superposition of hadronic matter, the strong interaction is 
suppressed and the collision becomes essentially a very clean electromagnetic
process. 

In \cite{nos20} it was argued that forward pions are very likely to be
produced by magnetic excitation (ME) of the nucleons in the nuclei. The   
strong classical magnetic  field produced by one nucleus induces magnetic
transitions, such as
$N \to \Delta$ (where $N$ is a proton or a neutron), in the 
nucleons of the other nucleus. The produced $\Delta$ keeps moving together
with the nucleus (or very close to it) and then decays almost 
exclusively through the reaction $\Delta \to N + \pi$. From the kinematics
we know that the pion has a very large longitudinal momentum and very
large rapidity. Since there is no other competing mechanism for forward pion
production in UPC's the observation of these pions would be a signature of  
the magnetic excitation of the nucleons and also an indirect measurement of 
the magnetic field.  In \cite{nos20} it was shown that ME has a very large
cross section. 

The hypothesis that in heavy ion collisions the electromagnetic field can be
treated classically and one can speak of a classical magnetic field has never
been tested. The classical field approximation may be expected to become a
reliable description of the quantum theory if the number of field quanta in each
field mode is sufficiently high. In this work we propose a way to test the
classical
approximation for the magnetic field.  To this end we consider again the
process discussed in \cite{nos20} 
\beq
\hskip6.5cm N \,  \to \,  \Delta \,  \to \,  N' \,  \pi
\label{transit}
 \eeq
but this time the transition is induced by photons
and not by the classical magnetic field.  We compute the same process using a
different formalism where the quanta of the field play the important role.
We then compare the results obtained with the two formalisms. In order to avoid
uncertainties associated with the spatial distribution of the nuclear matter
in the target we consider lead-proton collisions and choose to work in the
proton rest-frame.

In the next section we briefly review the formalism used in \cite{nos20} which
we shall call semi-classical and in the following section we describe the
quantum formalism, based on the equivalent photon approximation.
In the end  we compare the results obtained with the two methods.

\section{The semi-classical formalism}

A strong magnetic field can convert a hadron into another one with a different
spin, by ``flipping the constituent quark spins''. In ultra-relativistic 
heavy ion collisions this idea was first advanced in  \cite{muller1}, where
the authors studied the transition $\eta_c \to J/\psi$. In \cite{nos20} we
extended the calculations to the $N \to \Delta$ transition. 

\begin{figure*}[t]
\includegraphics[scale=0.30]{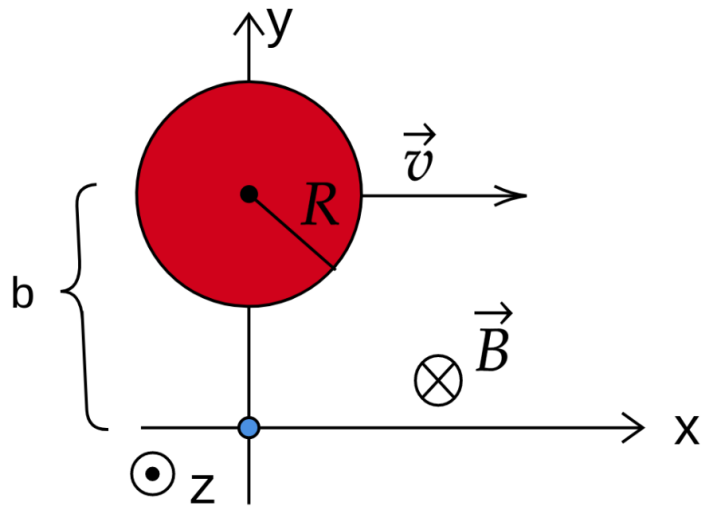} 
\caption{$Pb - p$ collision in the proton rest-frame. 
Coordinate system with magnetic field along the z direction. 
The projectile  nucleus of radius $R$ moves  with velocity $\vec{v}$    
and impact parameter $b $. The blue circle represents the proton at rest.}
\label{fig1}
\end{figure*}

Let us consider an ultra-peripheral $Pb - p$ collision, where the proton 
is at rest, as shown in Fig. \ref{fig1}. Under the influence of the strong
magnetic field generated by the moving nucleus,  the nucleon is converted into
a $\Delta$. For the sake of definiteness let us consider the 
transition $|p \uparrow \rangle  \to | \Delta^+ \uparrow \rangle$. 
The amplitude for this process is given by \cite{nos20,muller1}: 
\begin{equation}
    a_{fi} = -i \int_{-\infty}^{\infty} e^{iE_{fi}t'}
\langle \Delta^+ \uparrow| H_{int} (t') |p \uparrow \rangle \, dt'
\label{amp} 
\end{equation}
where $\hbar = 1$ and  $E_{fi}=(m_\Delta^2-m_n^2)/2m_n$,  where  
$m_{\Delta}$ and $m_n$ are the $\Delta$ and nucleon masses
respectively. The interaction Hamiltonian is given by:
\begin{equation}
H_{int} (t) = - \vec \mu . \vec B (t)
\label{hint}
\end{equation}
The magnetic dipole moment of the nucleon is given by the sum of the
magnetic dipole moments of the corresponding constituent quarks:
\begin{equation}
\vec \mu = \sum_{i=u,d}\vec \mu_i = \sum_{i=u,d} \frac{q_i}{m_i}\vec S_i
\label{mom}
\end{equation}
where $q_i$ and $m_i$ are the charge and constituent mass of the  quark of
type $i$ and
$\vec{S}_i$ is the spin operator acting on the spin state of  this quark. 
In Fig. \ref{fig1} we show the system of coordinates and the moving projectile.
The projectile of radius $R$  moves along the x direction with impact parameter
$b$ and the magnetic field is in the z direction. Since we are studying an UPC, 
we will, for simplicity, assume that the projectile-generated field is the
same produced by a point charge. The field is given by \cite{muller2}:
\begin{equation}
   B_z (t) = \frac{1}{4\pi}\frac{qv\gamma (b-y)}{((\gamma(x-vt))^2
        +(y-b)^2+z^2)^{3/2}}
\label{field}
\end{equation}
In the above expression $\gamma$ is the Lorentz factor, $b$ is the impact
parameter along the $y$ direction, $v \simeq 1$ is the 
projectile velocity and the projectile electric charge is $q =Z e$.

The interaction Hamiltonian acts on spin states. The relevant ones are:
\begin{equation}
|p \uparrow \ra
  =\frac{1}{3\sqrt{2}}[udu(\downarrow\uparrow\uparrow+\uparrow\uparrow\downarrow  
    -2\uparrow\downarrow\uparrow)+duu(\uparrow\downarrow\uparrow
    +\uparrow\uparrow\downarrow
    -2\downarrow\uparrow\uparrow)+uud(\uparrow\downarrow\uparrow
    +\downarrow\uparrow\uparrow
-2\uparrow\uparrow\downarrow)]
\end{equation}
\begin{equation}
|\Delta^+ \uparrow\ra=\frac{1}{3}(uud+udu+duu)(\uparrow\uparrow\downarrow
+\uparrow\downarrow\uparrow+\downarrow\uparrow\uparrow)
\end{equation}
With these ingredients we can compute  the matrix element 
$\la\Delta^+\uparrow|H_{int}|p \uparrow\ra$. 
It can be obtained by substituting Eqs. (\ref{mom})
and (\ref{field}) into Eq. (\ref{hint}) and then calculating the sandwiches
of $H_{int}$ with the spin states given above.
Evaluating  the  nucleon-delta transition matrix element we find:  
\beq
\la \Delta^+\uparrow|H_{int}|p \uparrow \ra \, = \,  \frac{\sqrt{2}e}{3m} B_z
\label{sand}
\eeq
The cross section for a single $N \to \Delta$ transition is given by:
\begin{equation}
    \sigma=\int |a_{fi}|^2 \, d^2b = 2\pi\, \int|a_{fi}|^2\, b \, db 
\end{equation}
where we have used cylindrical symmetry
$d^2b\, =\, b\, db\, d\theta \to 2\, \pi\,  b \, db$. Inserting  (\ref{sand}) 
into (\ref{amp}) and using it in the above expression we find:
\begin{equation}
  \sigma    =\frac{Z^2e^4}{9\pi m^2}
  \left( \frac{E_{fi}}{v\gamma} \right)^2 \int_{R}^{\infty}
\Big[ K_1\Big(\frac{E_{fi}b}{v\gamma}\Big)\Big]^2 b \, db
\label{sigclass}
\end{equation}
where $K_1$ is the modified Bessel function. This is the result obtained
with the
semi-classical  approach. For the purpose of this work it is enough to consider
a nucleon as a target. In \cite{nos20} we computed the cross section for a
nucleus-nucleus collision.

\section{The quantum formalism}
 
In the quantum formalism, the electromagnetic field produced by an
ultra-relativistic electric charge is replaced by a flux of photons
\cite{upc}.
\begin{figure}[th!]
  \centering
    \includegraphics[width=1.2\linewidth]{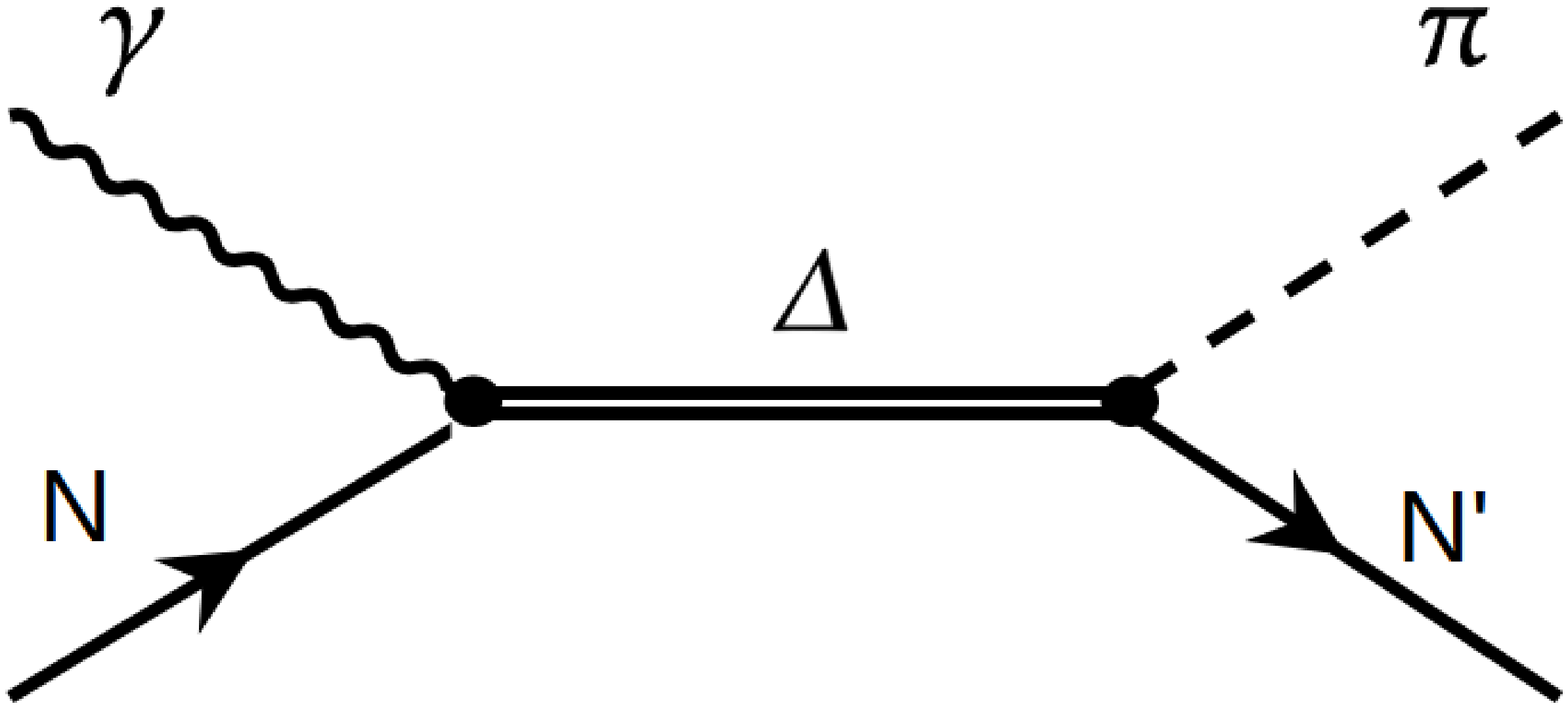}
    \caption{Quantum version of the process depicted in Fig. \ref{fig1}. 
        Pion photoproduction through a $\Delta$ resonance.}
    \label{feynman}
\end{figure}
Now, in a high energy UPC, the projectile becomes a source of almost real
photons and we replace the classical field by a collection of quanta. 
Thus, the cross section of the process (\ref{transit})  can be written in
a factorized form in terms of the photon flux produced by the projectile and
the photon-nucleon cross section \cite{upc}:
\begin{equation}
\sigma = \int \frac{d\omega}{\omega} \, n(\omega)
\,  \sigma_{\gamma N\rightarrow N\pi} (\omega)
\label{sigquan}
\end{equation}
In the above expression $n(\omega)$ represents the photon spectrum generated by
the source \cite{upc}:
\begin{equation}
  n(\omega) = \frac{Z^2 \alpha}{\pi}\Bigg[2\xi K_0(\xi)K_1(\xi)-\xi^2[K_1^2(\xi)
      -K_0^2(\xi)]\Bigg],  \hspace{2cm} \xi=\frac{\omega (R_1+R_2)}{\gamma}
\label{flux}
\end{equation}
where $\omega$ is the photon energy, $R_1$ and $R_2$ are the radii of the
projectile
and the target, parametrized as $R_A=1.2 \, A^{1/3}fm$, and $\gamma$ the
Lorentz boost in the target frame. From the above expression it is clear that
the average energy carried by an emitted photon increases with $\gamma$ and
hence with the collision energy $\sqrt{s}$. The photon average energy may be
estimated as 
\begin{equation}
  \bar{\omega}=
  \frac{\int_0^{\gamma m_n} \, n(\omega) \,  \omega \,  d \omega}
       {\int_0^{\gamma m_n} \, n(\omega)  \,  d \omega}
\label{average}
\end{equation}
In the LHC energy region $\gamma \simeq 1000$ and the above expression yields
$\bar{\omega} \simeq 10$ GeV.

In order to perform the calculation of the total cross section, it is necessary 
to know the cross section of the process
$\gamma N \rightarrow N \pi$. In a first approximation
$\sigma_{\gamma N \rightarrow N \pi}$ can be calculated evaluating the
Feynman diagram shown in Fig. \ref{feynman}.  This is a very well known 
process. In fact, there is an intense effort devoted to the study of  
nucleon resonances both experimentally and theoretically
\cite{ire20,pasca,clas,ab,manolo20,gil19}. Most of the interest lies on the
energy region around the threshold of $\Delta$ production, i.e.,
$\bar{\omega} \simeq 200 - 600$ MeV. As it was just mentioned above, we
are primarily interested in the high energy region, far from this threshold.
We need a
formula which correctly reproduces the behavior of the cross section in the 
$\Delta$ resonance region and which can be extrapolated to  higher energies.   
This is the most important source of uncertainty in the evaluation of
(\ref{sigquan}). 

\begin{figure}[th!]
    \centering
    \includegraphics[width=0.6\linewidth]{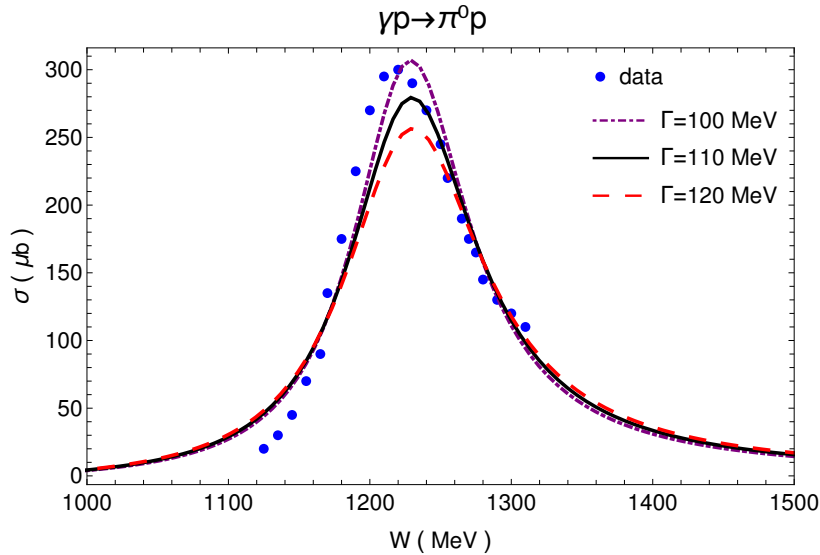}
    \caption{Pion photoproduction.
      Comparison between the theory, Eq. (\ref{js}), and data from
      \cite{datadel}. }
    \label{data}
\end{figure}

A simple parametrization of the $\pi^0$  photoproduction cross section can be
taken from Jones and Scadron \cite{Jones}:
\begin{equation}
  \sigma_{\gamma N \rightarrow N \pi} (\omega)
  =2\pi\int_0^\pi d\theta \, \sin{\theta} \,
  \frac{\alpha \, \omega }{12 \, m_n \,  W} \,
  \frac{\sin^2 \delta }{\Gamma} \, \left[ \, 
    |F_+^*|^2 \, f(\theta)\, + \, |G_+^*|^2 \, g(\theta)\, \right]
  \label{js}
\end{equation}
In the above expression $\alpha = 1/137 $, $W^2=m_n^2 + 2 \, \omega \, m_n$ is
the photon-nucleon center of mass energy squared,
$m_n$ is the nucleon mass, $\Gamma$ is the $\Delta$
decay width and $F_+^*=G_M^*-3G_E^*$, $G_+^*=G_M^*+G_E^*$. The form factors
$G_M^*$ and $G_E^*$ are functions of the photon virtuality $Q^2$.
Since we are interested in photoproduction they are taken at $Q^2=0$.  
The calculations were all carried out in the laboratory frame.
The angular dependence is given by $ f(\theta)= (3\cos^2\theta+1)/2$, 
$g(\theta)= (9 \sin^2 \theta)/2$ and 
\begin{equation}
\sin^2\delta=| \frac{m_\Delta \Gamma}{(W^2 - m_\Delta^2+im_\Delta\Gamma})|^2
\end{equation}
The  expression (\ref{js}) contains three parameters $G_M^*$, $G_E^*$  
and $\Gamma$, which can be determined by fitting the experimental data
on $\pi^0$ photoproduction. We have adjusted (\ref{js}) to the data
published in
\cite{datadel}.  The result is shown in Fig. \ref{data}. In order to estimate
the uncertainty in the extrapolation of (\ref{js}) to higher photon energies 
we have varied the $\Delta$ width within the interval $100 < \Gamma < 120$ MeV.
As it can be seen, the high energy tail of the curve is not very sensitive to
changes in $\Gamma$. The uncertainties in $G_M^*(0)$ and $G_E^*(0)$ are very
small
and changes of these quantities would not significantly change the cross
sections.

\begin{figure}
{\includegraphics[width=0.7\linewidth]{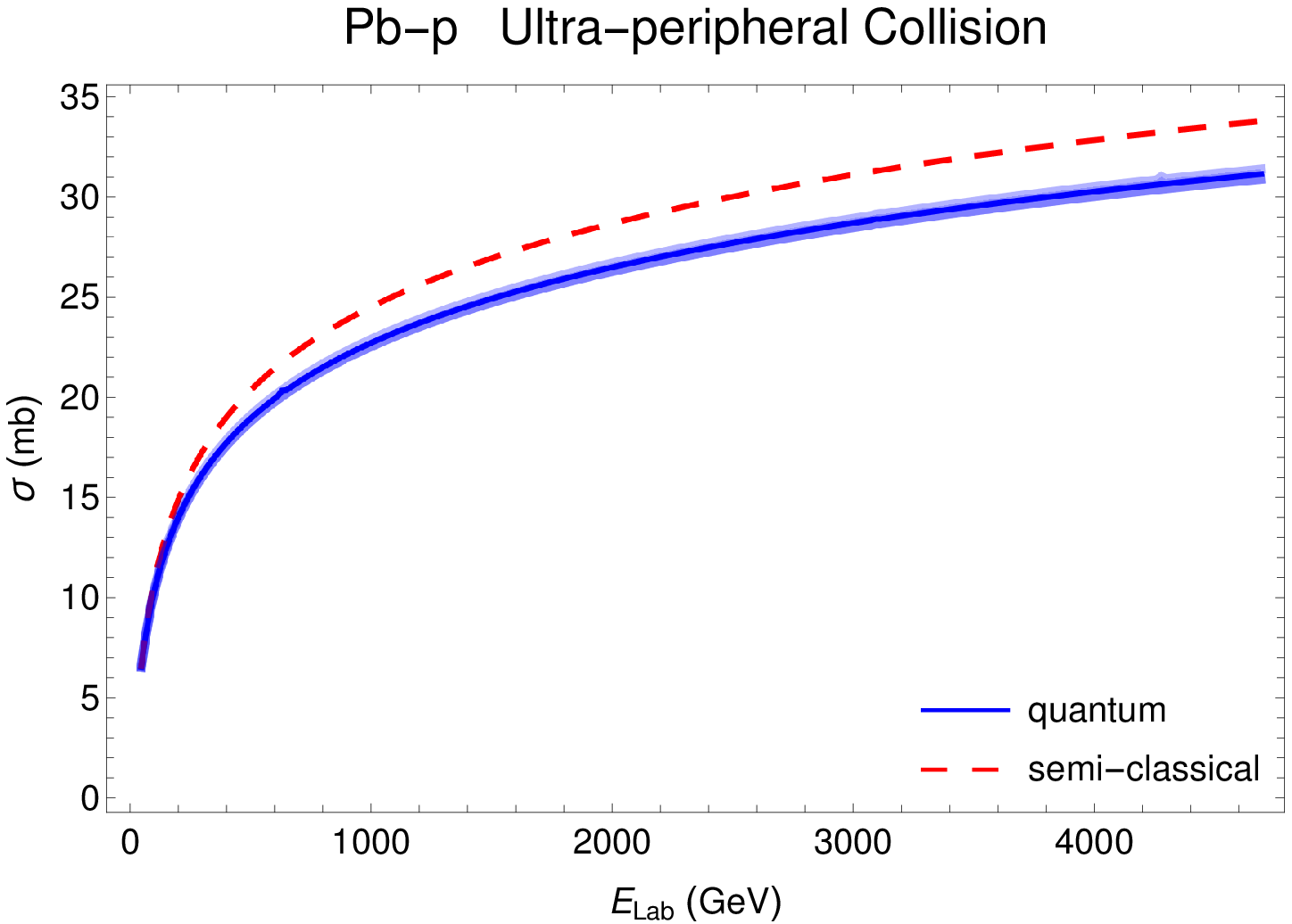}} \\
\vskip15mm
{\includegraphics[width=0.75\linewidth]{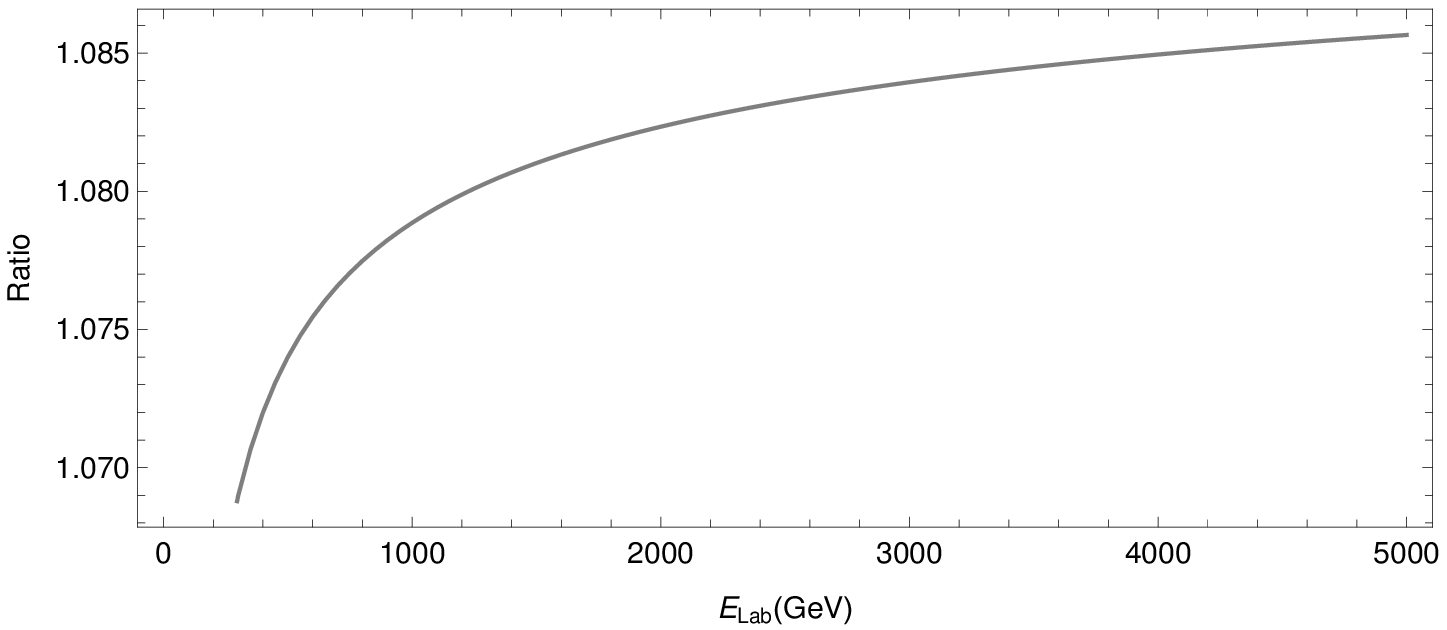}}
\caption{Comparison between the semi-classical and quantum cross sections  
  of pion photoproduction in an ultra-peripheral lead-nucleon collision.
  Upper pannel: cross sections as a function of the energy per nucleon
  in the laboratory frame.
  Lower pannel: ratio between semi-classical and quantum cross sections.}
    \label{comp}
\label{kappa}
\end{figure}

Having determined $\sigma_{\gamma N \rightarrow N \pi}$,  we insert it into
(\ref{sigquan}) and evaluate the cross section of the quantum process. The 
results are then compared with the results obtained with the semi-classical
approach (given by (\ref{sigclass})) and presented in Fig. \ref{comp}. 
The cross sections are plotted as a function of the energy per nucleon (of
the projectile) in the laboratory frame $E_{Lab} = \gamma \, m_n$. 
In the
upper pannel we compare the curves obtained with (\ref{sigclass})
(dashed line) and with (\ref{sigquan}) (solid lines).
The band in the lower curve represents
the different choices of the width $\Gamma$, i.e., the different values shown
in Fig. \ref{data}. In the lower pannel we show 
(\ref{sigclass}) divided by the central value of (\ref{sigquan}). This ratio
quantifies the difference between these two curves and it approaches 9 \%
at the highest energies. 
These results suggest that the classical approximation of
the magnetic field reproduces most of the photon interaction in
photoproduction in high energies.

The quantum formula (\ref{js}) could be improved.  At low energies
there are other resonances. To improve the accuracy of the extrapolation
to higher photon energies, it would be necessary to change (\ref{js}) including
higher resonances or, alternatively, define some procedure to ``average over the
bumps'', as it was done (although in a different context), for example, in
\cite{brona}.

\section{Concluding remarks}

In heavy ion collisions the sources are so intense that one can treat
classically the electromagnetic field. In particular, one can compute
the magnetic field and use it to make a number of predictions. Although
plausible, this conjecture had never been tested before. In this work we have
devised a test for this idea. We have found a process which can be
calculated in two different ways: one using the magnetic field and  one  
relying solely on quantum
physics. The EPA method has been extensively used and has yielded
predictions confirmed by experimental data. 

Our results give some support to the classical approximation for the magnetic
field and hence give support to all the calculations done previously based
on this approximation. 


\begin{acknowledgments}
The authors are deeply grateful to G. Ramalho for very instructive discussions.
This work was  partially financed by the Brazilian funding agencies CAPES and
CNPq.
\end{acknowledgments}



\begin{thebibliography}{99}

\bibitem{skokov}     V.~Skokov, A.~Y.~Illarionov and V.~Toneev,
                     Int.\ J.\ Mod.\ Phys.\ A {\bf 24}, 5925 (2009).


\bibitem{voro}       V.~Voronyuk, V.~D.~Toneev, W.~Cassing,
                     E.~L.~Bratkovskaya, V.~P.~Konchakovski and S.~A.~Voloshin, 
                     Phys.\ Rev.\ C {\bf 83}, 054911 (2011).


\bibitem{muller2}    M.~Asakawa, A.~Majumder and B.~Muller,
                     Phys.\ Rev.\ C {\bf 81}, 064912 (2010).   


\bibitem{hattori}    For a recent review see  K.~Hattori and X.~G.~Huang,
                     Nucl.\ Sci.\ Tech.\  {\bf 28}, 26 (2017) and references
                     therein;
                     A.~Dubla, U.~Gürsoy and R.~Snellings,
                     arXiv:2009.09727; 
                     C.~S.~Machado, F.~S.~Navarra, E.~G.~de Oliveira,
                     J.~Noronha and M.~Strickland,
                     Phys.\ Rev.\ D {\bf 88}, 034009 (2013). 
 

\bibitem{cme}        K.~Fukushima, D.~E.~Kharzeev and H.~J.~Warringa,
                     Phys.\ Rev.\ D {\bf 78}, 074033 (2008). 


\bibitem{upc}        C.~A.~Bertulani, S.~R.~Klein and J.~Nystrand,
                     Ann.\ Rev.\ Nucl.\ Part.\ Sci.\  {\bf 55}, 271 (2005);  
                     V.~P.~Goncalves and M.~V.~T.~Machado, 
                     Eur.\ Phys.\ J.\ C {\bf 40}, 519 (2005).


\bibitem{nos20}      I.~Danhoni and F.~S.~Navarra, 
                     Phys.\ Lett.\ B {\bf 805}, 135463 (2020). 

                     
\bibitem{muller1}    D.~L.~Yang and B.~Muller,
                     J.\ Phys.\ G {\bf 39}, 015007 (2012).  
                     
\bibitem{ire20}      For a very recent and comprehensive theoretical and
                     experimental  review, see
                     D.~G.~Ireland, E.~Pasyuk and I.~Strakovsky,
                     Prog.\ Part.\ Nucl.\ Phys.\  {\bf 111}, 103752 (2020)
                     and references therein. 
                     
\bibitem{pasca}      V.~Pascalutsa, M.~Vanderhaeghen and S.~N.~Yang,
                     Phys.\ Rept.\  {\bf 437}, 125 (2007). 

\bibitem{clas}       I.~G.~Aznauryan {\it et al.} [CLAS Collaboration],
                     Phys.\ Rev.\ C {\bf 80}, 055203 (2009). 

\bibitem{ab}         I.~G.~Aznauryan and V.~D.~Burkert,
                     Prog.\ Part.\ Nucl.\ Phys.\  {\bf 67}, 1 (2012). 

                     
\bibitem{manolo20}   G.~H.~G.~Navarro and M.~J.~Vicente Vacas,
                     arXiv:2008.04244 [hep-ph].
                     

\bibitem{gil19}      G.~Ramalho,
                     Phys.\ Rev.\ D {\bf 100}, 114014 (2019);
                     Eur.\ Phys.\ J.\ A {\bf 55}, 32 (2019);
                     Eur.\ Phys.\ J.\ A {\bf 54}, 75 (2018);
                     Phys.\ Rev.\ D {\bf 94}, 114001 (2016).

                     
\bibitem{Jones}      H.F. Jones and M.D. Scadron,
                     Annals of Phys. {\bf 81}, 1 (1973).


                     
\bibitem{datadel}    D.~A.~McPherson, D.~C.~Gates, R.~W.~Kenney and W.~P.~Swanson,
                     Phys.\ Rev.\  {\bf 136}, B1465 (1964);
                     M.~MacCormick {\it et al.}, 
                     Phys.\ Rev.\ C {\bf 53}, 41 (1996).

                     
\bibitem{brona}      S.~J.~Brodsky and F.~S.~Navarra, 
                     Phys.\ Lett.\ B {\bf 411}, 152 (1997). 


\end{thebibliography}
\end{document}